\documentclass[aps,prl,twocolumn,showpacs,superscriptaddress,nofootinbib,preprintnumbers]{revtex4-2}

\usepackage{graphicx}
\usepackage{physics}
\usepackage{amsmath}
\usepackage{amsthm}
\usepackage{slashed}
\usepackage[dvipsnames]{xcolor}
\usepackage{mathtools}
\usepackage{hyperref}
\usepackage[mathlines]{lineno}

\DeclareUnicodeCharacter{039B}{$\Lambda$}

\begingroup
\catcode`\&=\active
\gdef\EnableHTMLAmpersand{%
  \catcode`\&=\active
  \def&##1;{\&}%
}
\endgroup

\hypersetup{
    colorlinks = true
    ,urlcolor=blue
    ,anchorcolor=blue
    ,citecolor=blue
    ,filecolor=blue
    ,linkcolor=blue
    ,menucolor=blue
    ,linktocpage=true
    ,pdfproducer=medialab
}

\begin{document}

\title{Probing the Cosmic Axion Background via Axion-Photon Conversion in Filaments}

\author{Matthew J. Baldwin}
\email{mjbaldwin@uchicago.edu}
\affiliation{Department of Physics and KICP, University of Chicago, Chicago, IL 60637, USA}
\affiliation{Enrico Fermi Institute, Leinweber Institute for Theoretical Physics, University of Chicago, Chicago, IL 60637, USA}
\author{Gordan Krnjaic}
\email{krnjaicg@uchicago.edu}
\affiliation{Department of Astronomy \& Astrophysics and KICP, University of Chicago, Chicago, IL 60637}
\affiliation{Theoretical Physics Division, Fermi National Accelerator Laboratory, Batavia, IL 60510, USA}
\author{Duncan Rocha}%
 \email{drocha@uchicago.edu}
\affiliation{Department of Physics and KICP, University of Chicago, Chicago, IL 60637, USA}
\affiliation{Enrico Fermi Institute, Leinweber Institute for Theoretical Physics, University of Chicago, Chicago, IL 60637, USA}

\date{\today}

\begin{abstract}
The cosmic axion background (CaB) is a hypothetical population of relativistic axions produced in the early universe. If the CaB is produced from dark matter decays, the axions in this population can convert to photons in the magnetic fields of cosmological filaments, resulting in an isotropic gamma ray background flux. We present new indirect detection constraints on the axion mass and axion-photon coupling, for GeV-TeV dark matter with a decay lifetime below $10^{30}$ sec, by comparing this flux against experimental data. We exclude significant parameter space for axion masses below $10^{-5}$~eV for a broad range of dark matter masses and lifetimes, assuming conservative filament magnetic field strengths ($\sim 1$ nG). For large filament fields ($\sim 100$ nG), our strategy also constrains a portion of the QCD-axion parameter space for TeV-scale dark matter masses.

\end{abstract}

\preprint{ FERMILAB-PUB-26-0478-T}
\maketitle

\section{Introduction}\label{sec:level1}
Dark matter comprises a majority fraction of the universe's matter density. Although its gravitational effects are well characterized, the fundamental properties of dark matter remain mysterious. Direct detection searches probe the interaction of TeV-scale dark matter with Standard Model particles and place stringent bounds on nuclear scattering cross sections for a wide range of dark matter masses~\cite{LZCollaboration:2024lux,2tcc-bqck}. Many of the simplest extensions of the Standard Model containing dark matter candidates have been largely excluded by these increasingly sensitive searches. In light of this, it becomes well-motivated to consider more exotic interactions of TeV-scale dark matter with particles beyond the Standard Model. 

The axion-- whether the \emph{QCD axion} first proposed in the context of the strong CP problem~\cite{Peccei:1977hh,Peccei:1977ur,Weinberg:1978,Wilczek:1978}, or an \emph{axion-like particle} (ALP)~\cite{Marsh:2015xka}-- 
is one such particle that may couple to dark matter. 
If this coupling allows dark matter to decay to relativistic axions, such models predict the existence of a cosmic axion background (CaB), distinct from the dark matter population \cite{Dror:2021nyr,Langhoff:2022bij}. As CaB axions encounter diffuse magnetic fields, they can convert to photons through the Primakoff effect~\cite{Raffelt:1988,Sikivie1983}, and thereby contribute to the isotropic gamma ray background (IGRB). 

\begin{figure}
    \centering
    \includegraphics[width=\linewidth]{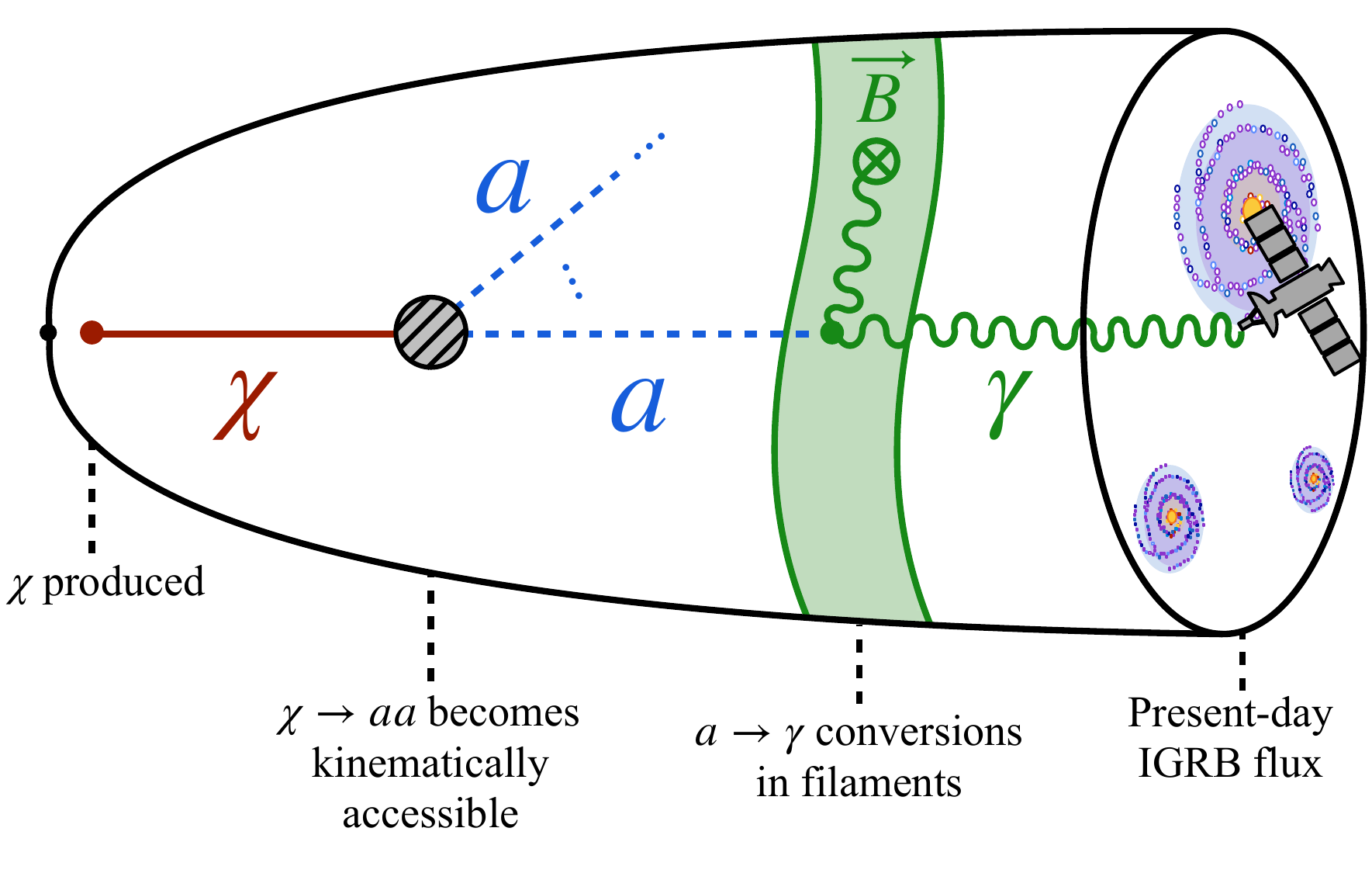}
    \caption{Schematic of the process that converts dark matter $\chi$ into an IGRB flux via axions that convert to photons in cosmological filaments.}
    \label{fig:sketch}
\end{figure}

Cosmological filaments provide a good candidate magnetic field within which axion-photon conversions may occur, as they fill a significant fraction of the universe with a non-negligible magnetic field~\cite{Amaral:2021,Vernstrom:2021,Carretti:2024bcf}. A schematic of this process is shown in Fig.~\ref{fig:sketch}.  Observing the IGRB produced from axion conversions in cosmological filaments offers a powerful indirect probe of dark matter properties.

In this paper, we constrain the axion mass and axion-photon coupling in the scenario where dark matter decays primarily to axions. We find that under this assumption, a significant range of dark matter decay lifetimes and
masses can be constrained from observations of the IGRB flux using the Fermi Large Area Telescope (LAT)~\cite{Fermi-LAT}, Energetic Gamma Ray Experiment Telescope (EGRET)~\cite{Strong_2004}, Compton Telescope (COMPTEL)~\cite{kappadath1998phd} and International Gamma-ray Astrophysics Laboratory (INTEGRAL)~\cite{Bouchet_2008}.

\begin{figure*}[ht]
    \centering
    \begin{minipage}[t]{0.49\textwidth}
        \centering
        \includegraphics[width=1\linewidth]{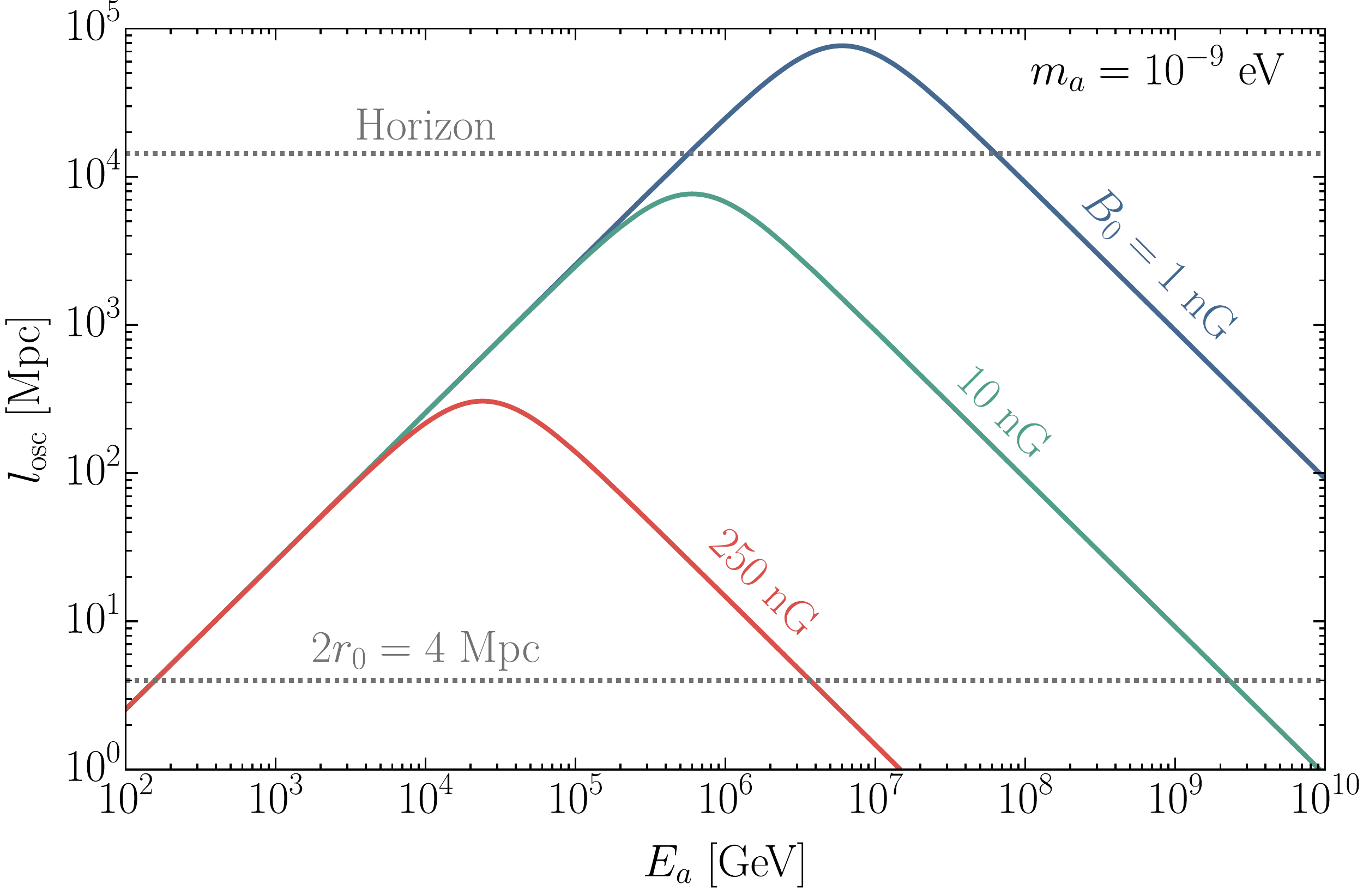}
    \end{minipage}%
    ~ 
    \begin{minipage}[t]{0.49\textwidth}
        \centering
        \includegraphics[width=1\linewidth]{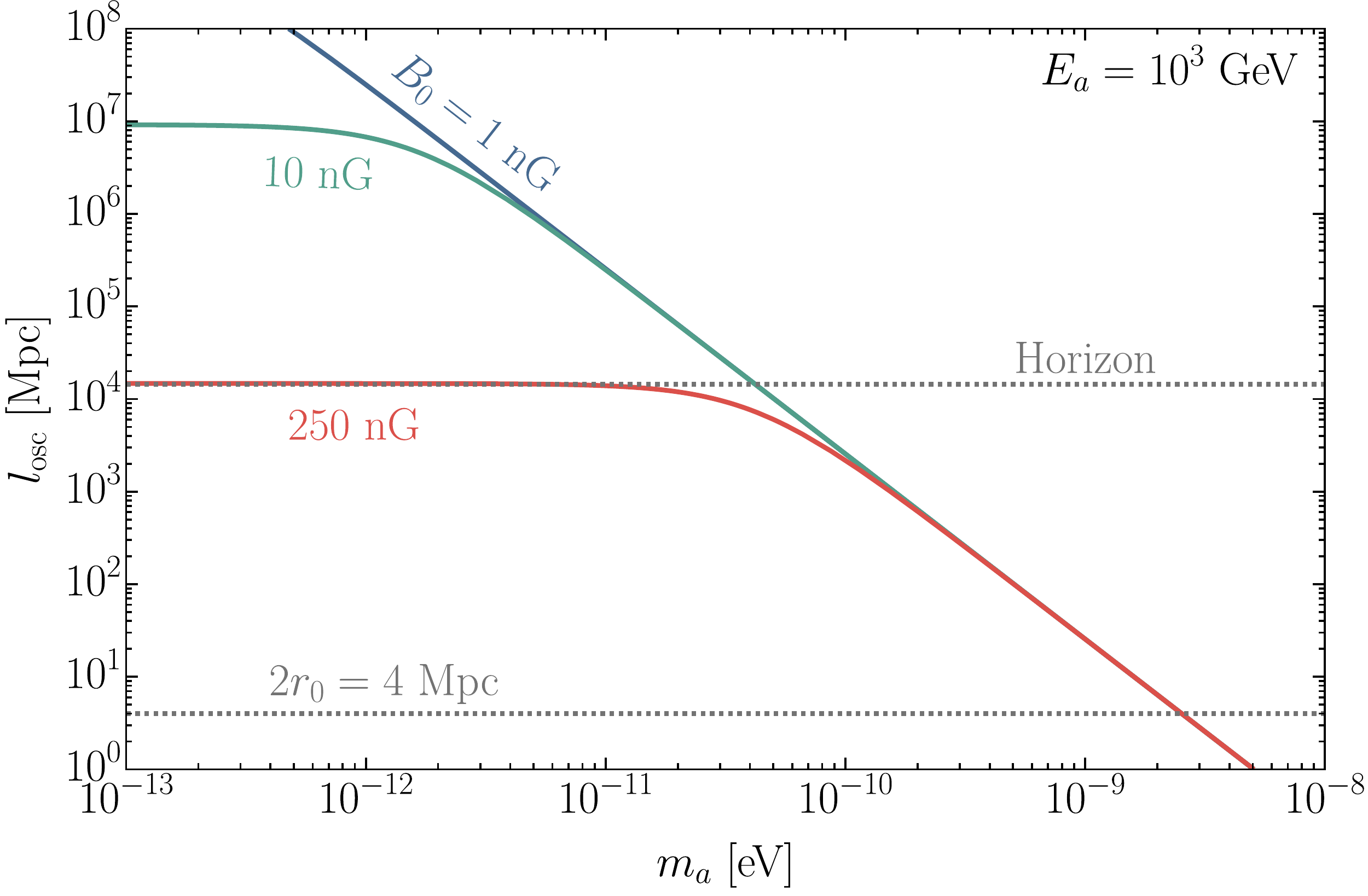}
    \end{minipage}
    \caption{Present-day oscillation lengths as a function of $E_a$ for a fixed $m_a$ (left), and $m_a$ for a fixed $E_a$ (right). The grey dashed lines, labeled ``Horizon'' and ``$2r_0=4$ Mpc'' show the present-day cosmological horizon and the average length an axion traverses through a filament, respectively. \textbf{Left:} to the right of the peak, the oscillation length is dominated by $m_\text{EH}$, and $l_\text{osc}\sim 1/E_a$; to the left of the peak, the oscillation length is dominated by either $m_\gamma$ or $m_a$, and $l_\text{osc} \sim E_a$. \textbf{Right:} as $m_a$ decreases, $m_{\text{EH}}\gg m_a,m_\gamma$ and $l_\text{osc}$ becomes independent of $m_a$.}
    \label{fig:losc}
\end{figure*}

Previous work has constrained
dark matter decays to gravitons through graviton-photon conversion in cosmic filaments via the Gertsenshtein effect~\cite{Dunsky:2025pvd}. The ADMX collaboration has also constrained dark matter decays to relativistic axions via axion-photon conversion in the laboratory \cite{Dror:2021nyr,ADMX:2023rsk}. 
Our work is the first to study relativistic axion conversion in filaments to probe decaying dark matter at the weak-scale.

In Sec.~\ref{sec:axion-photon-con}, we introduce axion-photon conversions in cosmological filaments and compute the conversion probability as a function of redshift. In Sec.~\ref{sec:flux}, we compute the present-day IGRB flux from axion-photon conversions. In Sec.~\ref{sec:lifetimes} we present our constraints on the axion mass and axion-photon coupling for benchmark dark matter masses, decay lifetimes and filament magnetic field strengths. We conclude our work in Sec.~\ref{sec:conclusion}.

\section{Axion-Photon Conversions in Cosmological Filaments}\label{sec:axion-photon-con}
\subsection{QCD Axions and Axion-Like Particles}
Axions are massive pseudo-Nambu Goldstone bosons that can couple to photons. In this work, we assume that axions $a$ couple to dark matter $\chi$, giving rise to a $\chi\rightarrow~aa$ decay channel with a $100\%$ branching fraction. Our results are straightforwardly modified for $1\rightarrow n$ decays or a non-unity branching fraction, we briefly comment on these modifications in Sec.~\ref{sec:flux}. We denote the dark matter mass as $m_\chi$ and the dark matter decay lifetime as $\tau_\chi$. The relevant IR Lagrangian terms are
\begin{equation}
    \mathcal{L}\supset\frac{1}{4}g_{a\gamma\gamma}a F_{\mu\nu}\Tilde{F}^{\mu\nu}-\frac{1}{2}m_a^2 a^2+\mathcal{L}_{m_\chi}+\mathcal{L}_{\chi\rightarrow a a},
\end{equation}
where $g_{a\gamma\gamma}$ and $m_a$ are the axion-photon coupling~\footnote{To avoid sign ambiguities in $g_{a\gamma\gamma}$ due to the freedom to redefine $a\rightarrow-a$, we present results for $\abs{g_{a\gamma\gamma}}$.} and axion mass, respectively; $\mathcal{L}_{m_\chi}$ represents the suitable mass term for $\chi$; and $\mathcal{L}_{\chi\rightarrow a a}$ represents any terms that give rise to $\chi\rightarrow a a$ decays. In Appendix.~\ref{app:models}, we discuss models that can give rise to dark matter decays to axions, however the results presented in this paper are agnostic to any choice of model~\footnote{Our results will not depend on the specific representation of $\chi$, we only assume that $\chi$ is in a representation for which decays to two axions are allowed.}.

For ALPs, $g_{a\gamma\gamma}$ and $m_a$ are treated independently, whereas for the QCD axion,
\begin{equation}
    g_{a\gamma\gamma}\approx \left(0.203 \frac{E}{N}-0.39\right)\left(\frac{m_a}{\text{GeV}}\right)\text{GeV}^{-1},
\end{equation}
where $E/N$ is the UV model-dependent ratio of the Peccei-Quinn (PQ) symmetry electromagnetic and color anomalies, $E$ and $N$, respectively. Two common choices are the KSVZ axion~\cite{PhysRevLett.43.103,SHIFMAN1980493} with $E/N=0$, and the DFSZ axion~\cite{DINE1981199,Zhitnitsky:1980tq} with $E/N=8/3$.

\begin{figure*}[ht]
    \centering
    \begin{minipage}[t]{0.49\textwidth}
        \centering
        \includegraphics[width=1\linewidth]{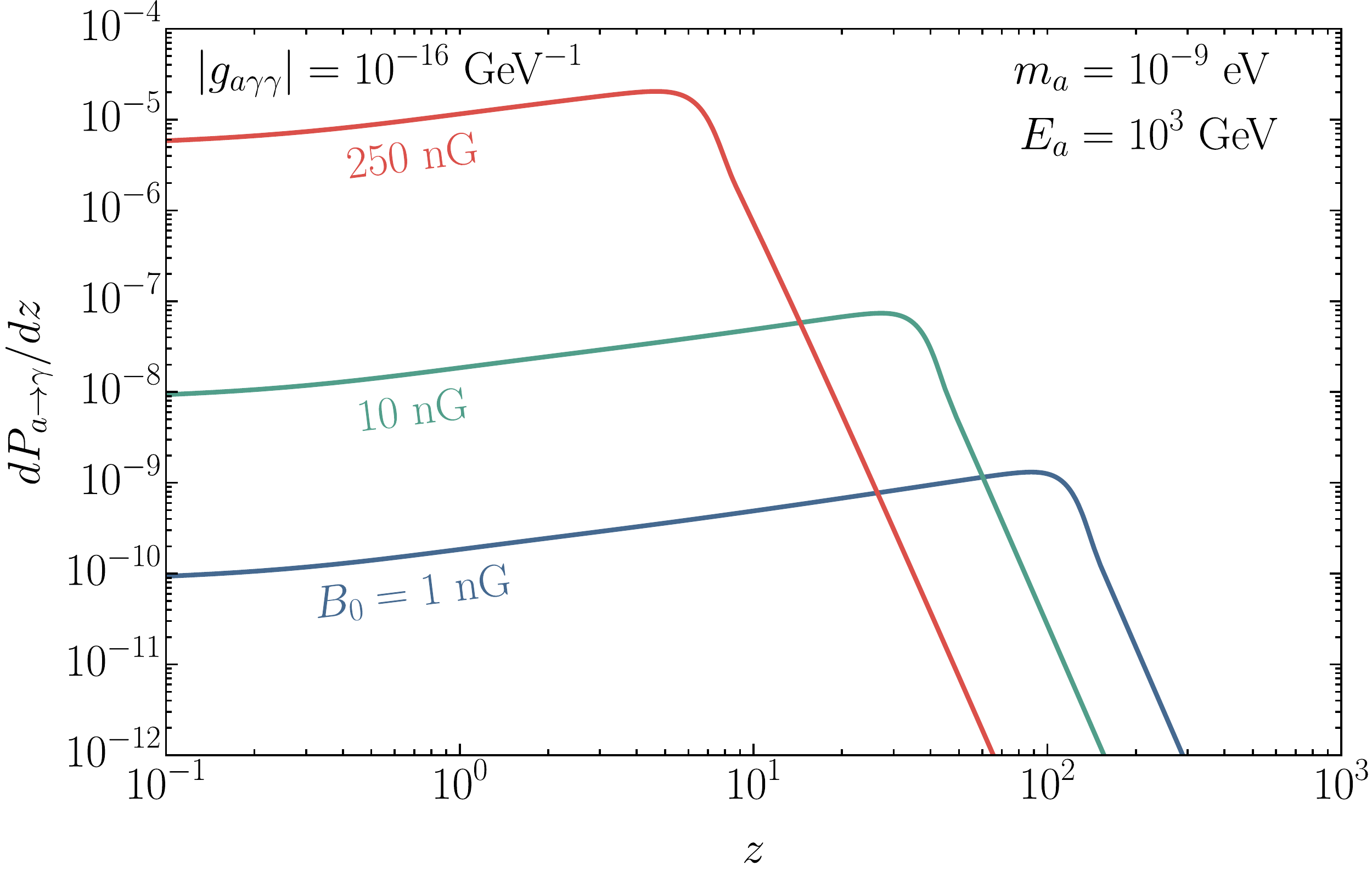}
    \end{minipage}%
    ~ 
    \begin{minipage}[t]{0.49\textwidth}
        \centering
        \includegraphics[width=1\linewidth]{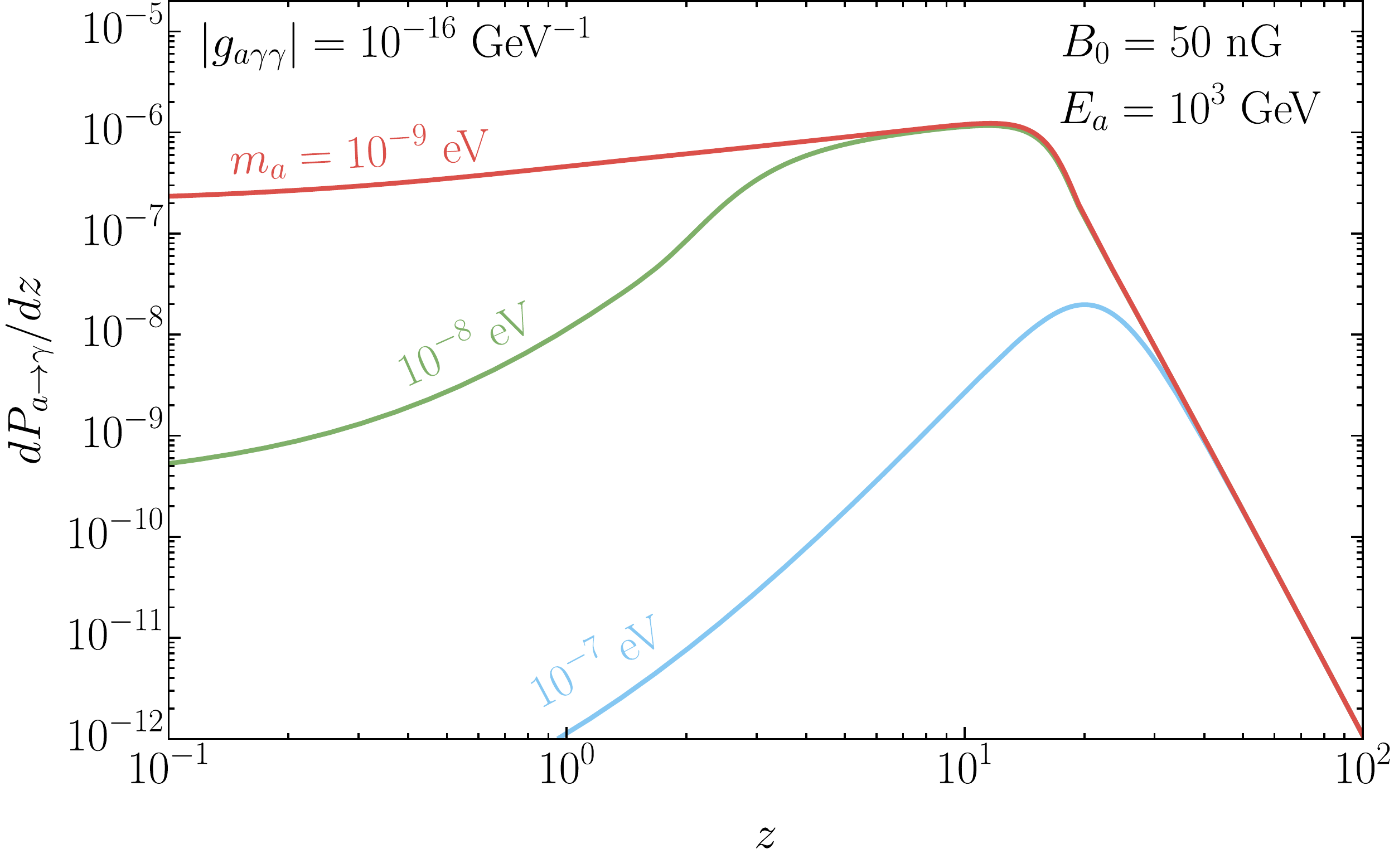}
    \end{minipage}
    \caption{Differential probability per unit redshift for an axion with $g_{a\gamma\gamma}=10^{-16}\text{ GeV}^{-1}$ and $E_a=10^3$ GeV for: a range of magnetic field strengths with $m_a=10^{-9}$ eV (left); and a range of axion masses with $B_0=50$ nG (right).}
    \label{fig:dPdzaverage}
\end{figure*}

\subsection{The Primakoff Effect}
The axion-photon conversion probability at redshift $z$ for an axion that traverses a length $l_f(z)$ within a background magnetic field $B(z)$ is~\cite{Marsh:2015xka}
\begin{equation}
    P^{(1)}_{a\rightarrow\gamma}(z)=\frac{g_{a\gamma\gamma}^2B^2(z) }{4}l^2_\text{osc}(z)\sin^2\left[\frac{ \, l_f(z)}{l_{\text{osc}}(z)}\right],
\end{equation}
where we have defined an \emph{oscillation length}, $l_{\text{osc}}(z)$, as
\begin{equation}
    l_{\text{osc}}(z)=\frac{4E_a(1+z)}{\abs{m_\gamma^2(z)-m_{\text{EH}}^2(z)-m_a^2}},
\end{equation}
where $E_a$ is the present-day ($z=0$) axion energy, and $m_\gamma(z)$ and $m_{\text{EH}}(z)$ are the photon plasma and Euler-Heisenberg masses, respectively, given by
\begin{equation}
\begin{split}
    &m_\gamma(z)=10^{-14}~(1+z)^{\frac{3}{2}}\left(\frac{n_e}{10^{-6}\text{cm}^{-3}}\right)^{\frac{1}{2}}\text{ eV},\\
    &m_{\text{EH}}(z)=\left(\frac{7 \alpha}{45 \pi}\right)^{\frac{1}{2}}\frac{B(z)}{B_c}E_a(1+z),
\end{split}
\end{equation}
where $n_e$ is the present-day ambient electron number density, and $B_c\approx m_e^2/e$.

\subsection{Cosmological Filaments}
Axion-photon conversions can occur within cosmological filaments, which fill an $\mathcal{O}(10\%)$ fraction of the universe with diffuse magnetic fields of $\mathcal{O}(1-100)$ nG~\cite{Amaral:2021,Vernstrom:2021,Carretti:2024bcf,Brown_2017,Vacca_2018,O_Sullivan_2019,Vernstrom_2019,O_Sullivan_2020,Locatelli_2021,Carretti_2022,Hoang_2023,anderson2024probingmagnetisedgasdistribution}. The magnetic field of a cosmological filament can be parameterized as~\cite{GRASSO2001163}
\begin{equation}
    B(z)=B_0(1+z)^2,
\end{equation}
where $B_0=\mathcal{O}(1-100)$ nG. The average path length of an axion through a filament and the second moment of the path length are~\cite{Dirac1943,Case1953,SANCHEZ20042211}
\begin{equation}
    \langle l_f(z)\rangle\approx\frac{2r_0}{1+z},~~\langle l_f^2(z)\rangle\approx \frac{16r_0^2}{3(1+z)^2},
\end{equation}
respectively, where $r_0\approx 2$ Mpc is the average filament radius determined from simulation~\cite{Colberg:2005,Wang:2024}.  In Fig.~\ref{fig:losc}, we show the present-day oscillation length $l_\text{osc}$ for three values of $B_0$, as a function of $E_a$ (left panel) and $m_a$ (right panel). In both plots we also show the cosmological horizon and average filament diameter. When $E_a$ is sufficiently small, $m_\text{EH}(z)$ can be neglected so that the denominator in $l_\text{osc}$ is dominated by $m_a$ or $m_\gamma(z)$ (whichever is larger), leading to $l_\text{osc}\propto E_a$. When $E_a$ is sufficiently large, $m_a$ and $m_\gamma(z)$ can be neglected~\footnote{Although $m_\gamma(z)$ is $z$-dependent, the values of $B_0$ and $E_a$ for which $m_\gamma(z)$ becomes larger than $m_\text{EH}(z)$ are never relevant in our analysis.} so that the denominator in $l_\text{osc}$ is dominated by $m_\text{EH}(z)$, leading to $l_\text{osc}\propto 1/E_a$.

\begin{figure*}[ht]
    \centering
    \begin{minipage}[t]{0.49\textwidth}
        \centering
        \includegraphics[width=1\linewidth]{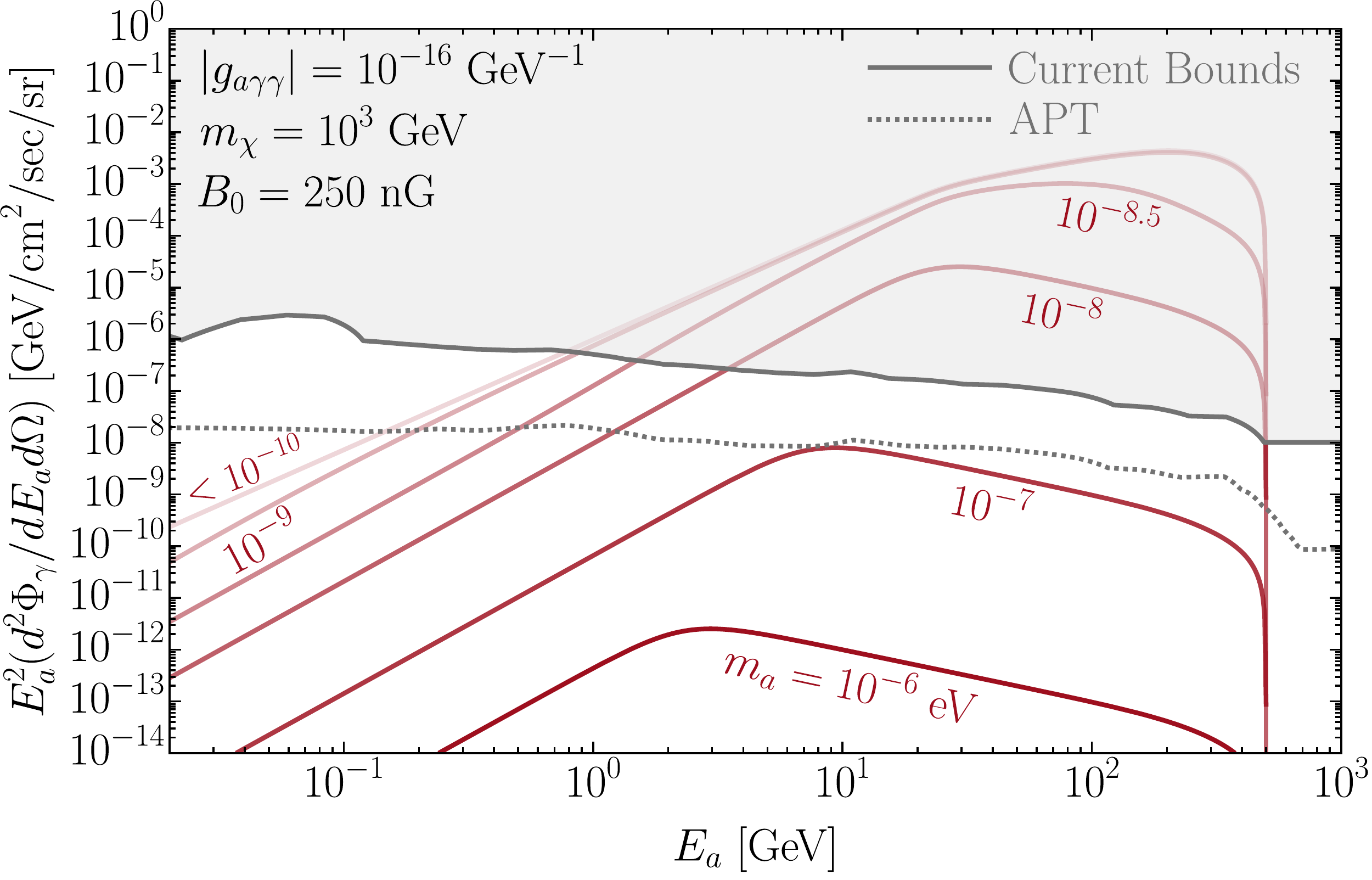}
    \end{minipage}%
    ~ 
    \begin{minipage}[t]{0.49\textwidth}
        \centering
        \includegraphics[width=1\linewidth]{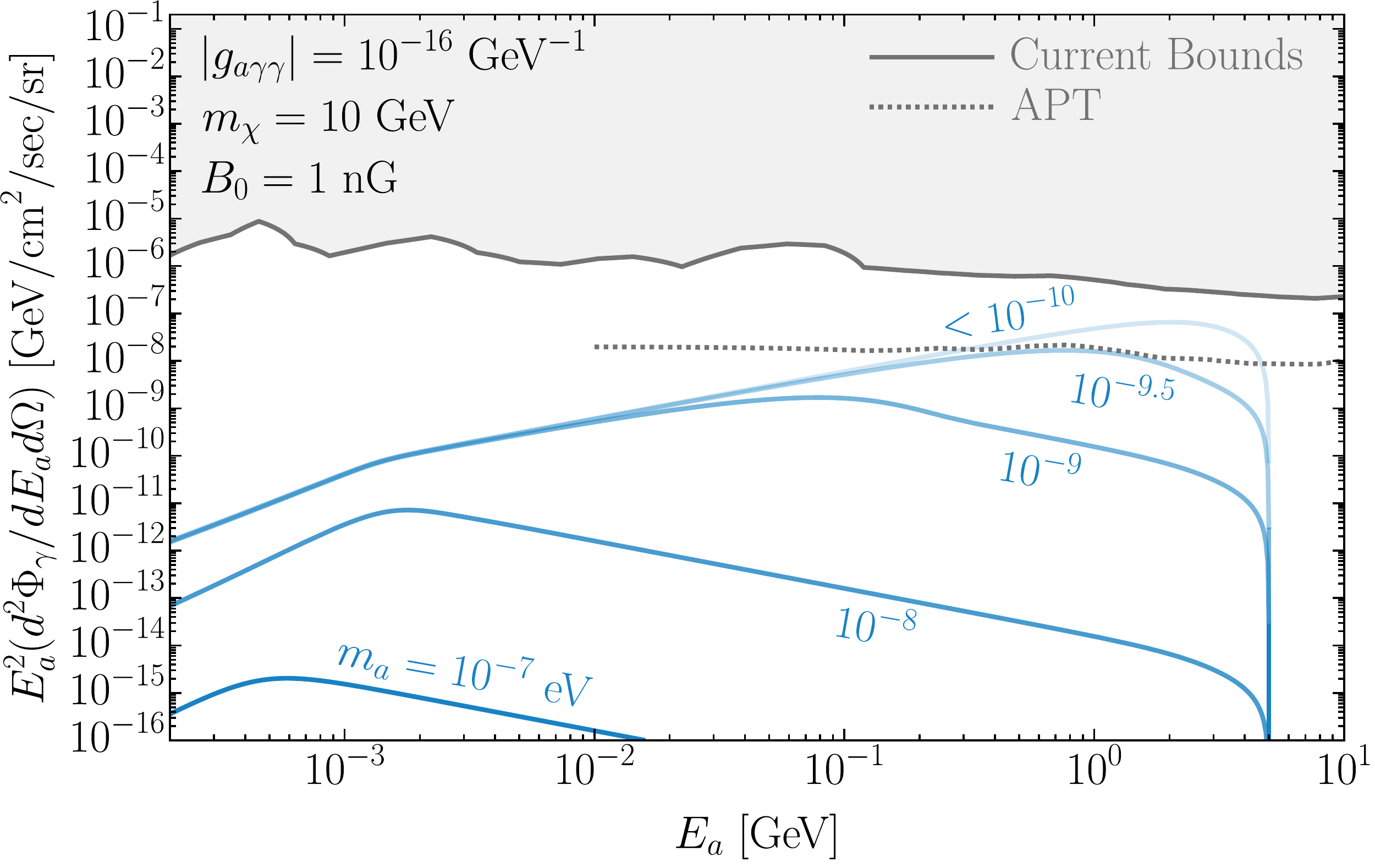}
    \end{minipage}
    \caption{
    Differential IGRB fluxes from axion-photon conversions within cosmological filaments for a dark matter lifetime $\tau_\chi=10^{19}$~sec, with: $m_\chi=10^3$ GeV and $B_0=250$~nG (left), and $m_\chi=10$ GeV and $B_0=1$~nG (right).  Note that as the axion mass is decreased, the curves in both panels converge, which accounts for the flattening of the constraints with respect to $m_a$ in Fig. \ref{fig:constraints}. 
    }
    \label{fig:fluxes}
\end{figure*}

When an axion traverses a cosmological filament, we model $l_f(z)$ as a Gaussian distributed variable with average $\langle l_f(z)\rangle$ and deviation $\sigma^2=\langle l_f^2(z)\rangle-\langle l_f(z)\rangle^2$, so that the axion-photon conversion probability, averaged over filament diameter, is
\begin{equation}
    \bar{P}^{(1)}_{a\rightarrow\gamma}(z)=\int_{0}^\infty dl_f\frac{P^{(1)}_{a\rightarrow\gamma}(z)}{\sqrt{2\pi\sigma^2}} \exp\left(-\frac{\left(l_f-\langle l_f\rangle\right)^2}{2\sigma^2} \right) .
\end{equation}
The integral can be computed analytically by the approximation of extending the domain of integration to $l_f\in(-\infty,\infty)$~\footnote{Extending the lower integral bound to $-\infty$ picks up $\approx15$\% error from the tail of the Gaussian distribution.},  for which we obtain
\begin{equation}\label{eq:Pavg}
\begin{split}
    \bar{P}^{(1)}_{a\rightarrow\gamma}(z)&\approx\frac{g_{a\gamma\gamma}^2B^2(z) l^2_\text{osc}(z)}{8}\\
    &\times\left[1-\exp(-\frac{2\sigma^2}{l_\text{osc}^2(z)})\cos(\frac{2\langle l_f\rangle}{l_\text{osc}(z)})\right].
\end{split}
\end{equation}
To understand the behavior of the IGRB flux that we will compute in Sec.~\ref{sec:flux} and the resulting axion constraints we present in Sec.~\ref{sec:lifetimes}, there are two relevant limits:
\begin{equation}\label{eq:Pav1scaling}
    \bar{P}^{(1)}_{a\rightarrow\gamma}(z)\approx\begin{cases}
        \frac{g_{a\gamma\gamma}^2B^2(z) \langle l_f\rangle^2}{8},&l_\text{osc}\gg \langle l_f\rangle,\\
        \frac{g_{a\gamma\gamma}^2B^2(z) l^2_\text{osc}(z)}{8},&l_\text{osc}\ll \langle l_f\rangle.
    \end{cases}
\end{equation}
When the oscillation length is much larger than the average length an axion traverses in a filament ($l_\text{osc}\gg \langle l_f\rangle$), the conversion probability becomes independent of $E_a$. In this limit, the conversion probability is enhanced.
When the oscillation length is much smaller than the average length an axion traverses in a filament ($l_\text{osc}\ll \langle l_f\rangle$), the conversion probability will either scale as $E_a^2$ (when $l_\text{osc}$ is $m_a$ or $m_\gamma$ dominated) or $1/E_a^2$ (when $l_\text{osc}$ is $m_\text{EH}$ dominated). In this limit, the conversion probability is suppressed.

Let us now compute the cumulative probability of an axion-photon conversion occurring between redshifts $z_1$ and $z_2$. Over a small physical distance $dr$, an axion traverses a number of filaments
\begin{equation}
    dN_f=\frac{f_{\text{vol}}(z)}{\langle l_f(z)\rangle}dr=\frac{f_{\text{vol}}(z)}{\langle l_f(z)\rangle}\frac{dz}{H(z)(1+z)},
\end{equation}
where $f_{\text{vol}}(z)\approx0.15$ is the volume filling fraction of filaments at redshift $z$~\footnote{We assume an approximately constant volume filling fraction across redshift motivated by~\cite{Dome_2023,Libeskind_2017}.}, and
\begin{equation}
    H(z)=H_0\sqrt{\Omega_m(1+z)^3+\Omega_\Lambda},
\end{equation}
is the Hubble parameter with $\Omega_m=0.31$, $\Omega_\Lambda=0.69$ and $H_0=68$ km/s/Mpc. The cumulative probability of an axion-photon conversion between $z_1$ and $z_2$ is then
\begin{equation}
\begin{split}
    P_{a\rightarrow\gamma}&=\int_{z_1}^{z_2} dN_f(z) \bar{P}^{(1)}_{a\rightarrow\gamma}\\
    &=\int_{z_1}^{z_2}dz~ \frac{f_{\text{vol}}(z)}{\langle l_f(z)\rangle}\frac{\bar{P}^{(1)}_{a\rightarrow\gamma}(z)}{H(z)(1+z)},
\end{split}
\end{equation}
from which we obtain the differential probability per unit redshift,
\begin{equation}
\begin{split}\label{eq:dPdz}
    \frac{dP_{a\rightarrow\gamma}}{dz}&=\frac{f_{\text{vol}}(z)}{\langle l_f(z)\rangle}\frac{\bar{P}^{(1)}_{a\rightarrow\gamma}(z)}{H(z)(1+z)}.
\end{split}
\end{equation}
In the left panel of Fig.~\ref{fig:dPdzaverage}, we show $dP_{a\rightarrow\gamma}/dz$ for an axion at a fixed $g_{a\gamma\gamma}$, $m_a$ and $E_a$, for three values of $B_0$. As $z$ increases, $l_\text{osc}$ decreases until it becomes smaller than $\langle l_f\rangle$. In this limit, $\bar{P}^{(1)}_{a\rightarrow\gamma}$ scales as $l_{\text{osc}}^2$ (see Eq.~(\ref{eq:Pav1scaling})). For $z\gg 1$, we have $l_{\text{osc}}^2\propto (1+z)^{-5}$ and $dP_{a\rightarrow\gamma}/dz$ quickly drops off. In the right panel of Fig.~\ref{fig:dPdzaverage}, we show $dP_{a\rightarrow\gamma}/dz$, at fixed $g_{a\gamma\gamma}$, $B_0$ and $E_a$, for three values of $m_a$. As $m_a$ decreases, $l_\text{osc}$ increases until $l_\text{osc}\gg \langle l_f\rangle$ (see Fig.~\ref{fig:losc}). In this limit, $dP_{a\rightarrow\gamma}/dz$ becomes independent of $m_a$~\footnote{In this case, our results are the same as those in Ref.~\cite{Dunsky:2025pvd} with the relevant substitution of parameters.}. Note-- as $g_{a\gamma\gamma}$ changes, for fixed $m_a$, $B_0$ and $E_a$, $dP_{a\rightarrow\gamma}/dz$ simply scales by $g_{a\gamma\gamma}^2$ for all $z$. In summary-- lighter axions with stronger axion-photon couplings have a higher probability of converting to photons~\footnote{To avoid the unphysical behavior in the parameter region for which $P_{a\rightarrow\gamma}> 1$, we modify as follows. Since the photon to axion conversion occurs with the same probability as the axion to photon conversion, we expect that for sufficiently large conversion probabilities, an equilibrium will be reached for which the number densities of axions and photons become equal. To account for this, we take the cumulative conversion probability to be $\min(P_{a\rightarrow\gamma},~0.5)$.}.

\section{Photon Flux}\label{sec:flux}
In this section, we compute the IGRB flux produced from axion-photon conversions in cosmological filaments, following dark matter decays.

\begin{figure*}[ht!]
    \centering
    \includegraphics[width=0.495\linewidth]{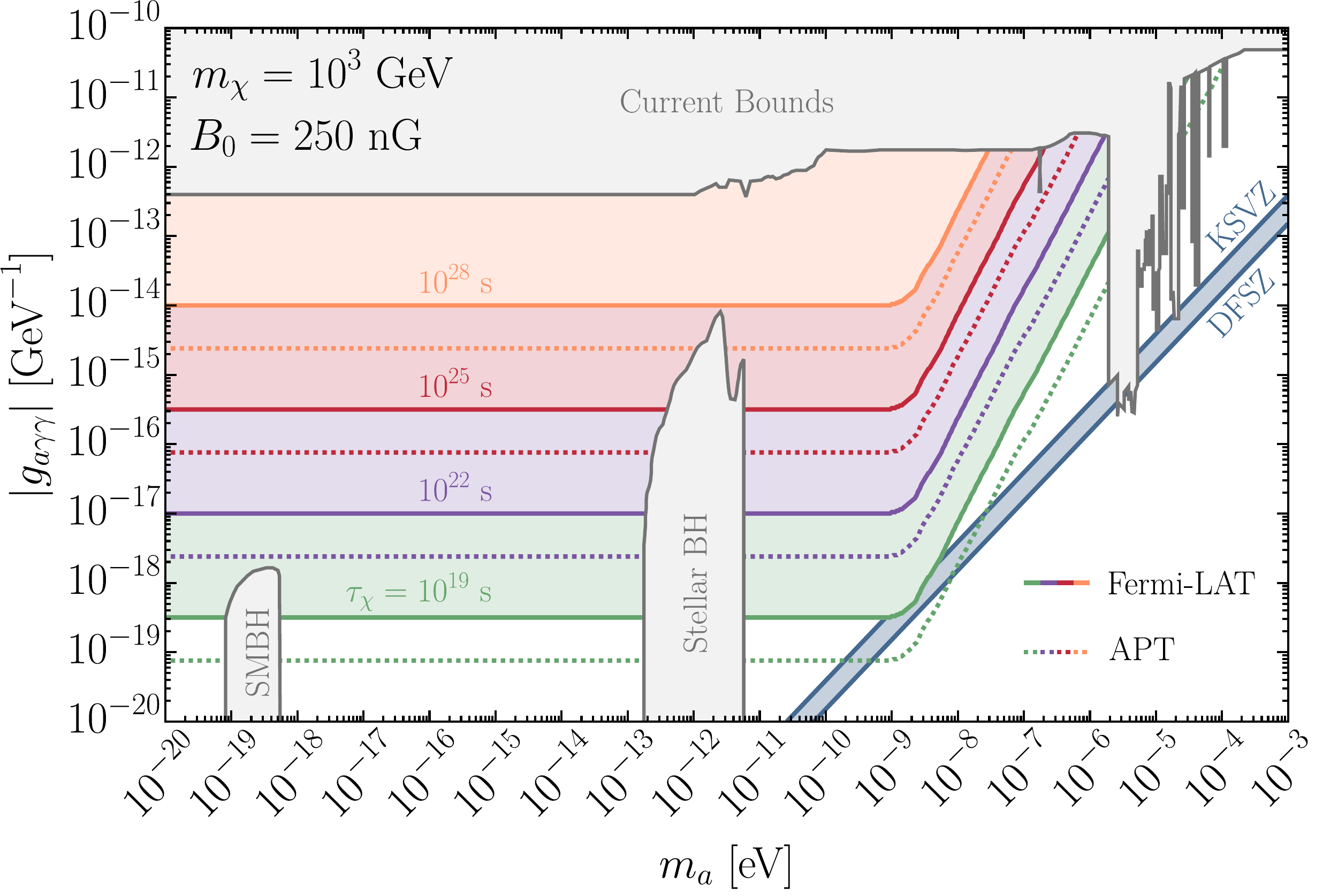}
    \includegraphics[width=0.495\linewidth]{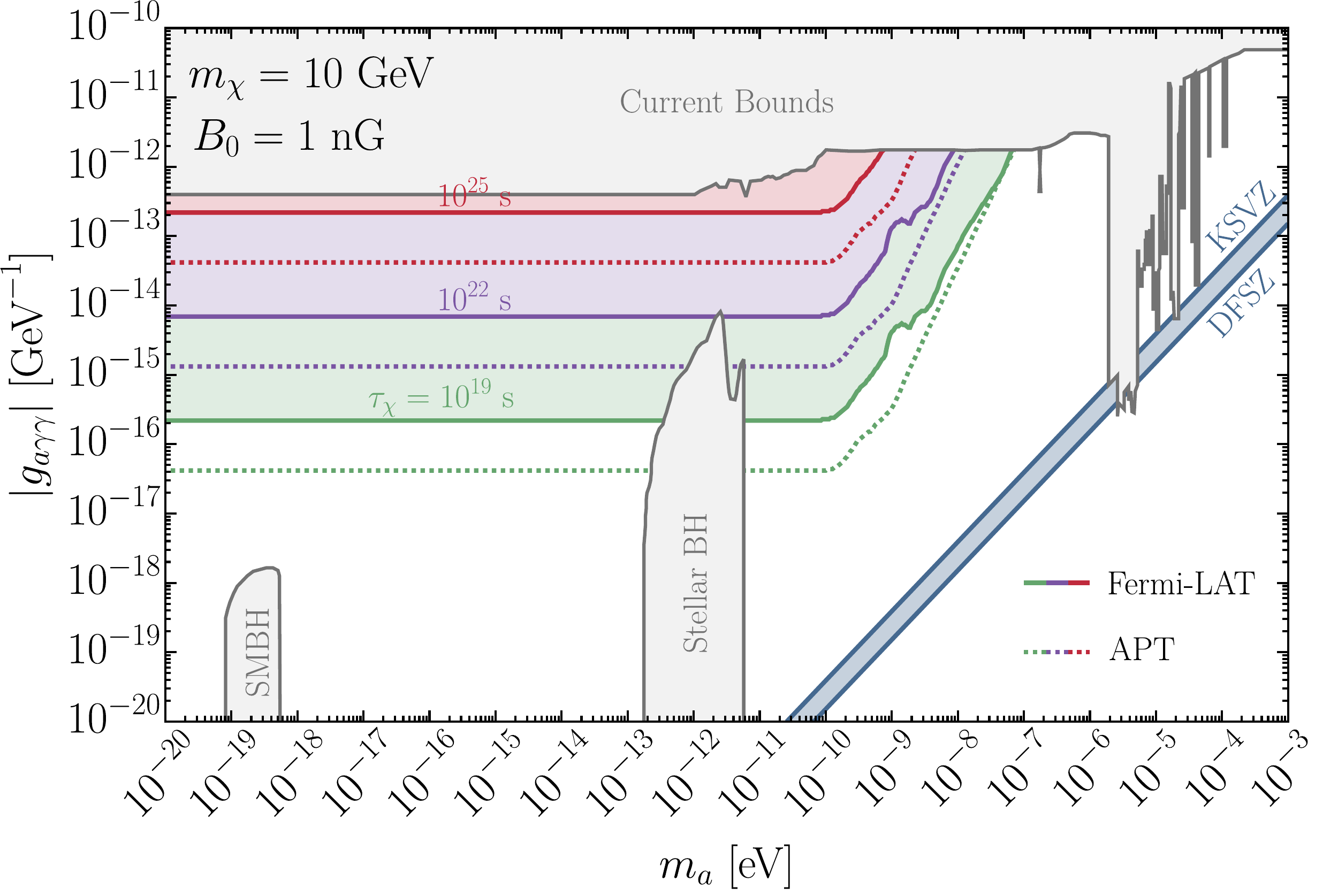}
        \caption{Constraints on the axion mass and axion-photon coupling using IGRB flux constraints, for various dark matter lifetimes and benchmark values $m_\chi=10^3$~GeV and $B_0=250$~nG (left), and $m_\chi=10$~GeV and $B_0=1$~nG (right). The $\tau_\chi=10^{19}$ sec constraint corresponds to the lifetime constraint set by $\Lambda$CDM LSS. Our constraints improve on current bounds, adapted from~\cite{AxionLimits}, for these benchmark parameters up to dark matter lifetimes of $\tau_\chi\approx10^{30}$ sec (left) and $\tau_\chi\approx10^{25}$ sec (right).}
    \label{fig:constraints}
\end{figure*}

\subsection{Axion Flux from Dark Matter Decays}
The flux of axions from dark matter decays is
\begin{equation}\label{eq:axion_flux_from_DM}
    \frac{d\Phi_a}{dE_a}(z_c)=\frac{\Omega_\chi\rho_c}{\tau_\chi m_\chi}\int_{z_c}^\infty dz\frac{1}{H(z)}\frac{d N_a}{dE_a}(z),
\end{equation}
where $\Omega_\chi=0.24$ is the present-day dark matter fraction, $\rho_c$ is the critical energy density, $z_c$ is the redshift at which axion-photon conversion in filaments begins, and
\begin{equation}\label{eq:specaxion}
    \frac{d N_a}{dE_a}(z)=2\delta\left(E_a(1+z)-\frac{1}{2}m_\chi\right)
\end{equation}
is the present-day axion energy spectrum due to dark matter decays that occurred at redshift $z$. Although we present results for a $1\rightarrow2$ decays with unity branching fraction, our results straightforwardly generalize-- by suitable modification of Eq.~(\ref{eq:axion_flux_from_DM}) and Eq.~(\ref{eq:specaxion})-- to $1\rightarrow n$ decays with a non-unity branching fraction.

\subsection{Photon Flux from Axion-Photon Conversions}\label{sec:IGRB}
The IGRB flux from axion-photon conversions within cosmological filaments is the convolution of the differential probability per unit redshift, given in Eq.~(\ref{eq:dPdz}), with the flux of axions from dark matter decays, given in Eq.~(\ref{eq:axion_flux_from_DM}). The IGRB flux is therefore
\begin{equation}\label{eg:photon_flux}
\begin{split}
    \frac{d\Phi_\gamma}{dE_a}(z)&=\int_z^\infty dz_c \frac{d\Phi_a}{dE_a}(z_c)\frac{dP_{a\rightarrow\gamma}}{dz_c}(z_c)\\
    &=\frac{2\Omega_\chi \rho_c}{\tau_\chi m_\chi E_a H(z_\text{max})}\int_z^{z_\text{max}} dz_c\frac{dP_{a\rightarrow\gamma}}{dz_c}(z_c),
\end{split}
\end{equation}
where $z_\text{max}=m_\chi/(2E_a)-1$ is the maximum redshift for which decays are kinematically accessible. Fermi-LAT places leading constraints on the IGRB flux for $E_a$ in the 1~GeV--~1~TeV range. These constraints will be further improved by the Advanced Particle-astrophysics Telescope (APT)~\cite{Alnussirat:2021/6}. For sub-GeV IGRB flux, we use the leading constraints from EGRET, COMPTEL and INTEGRAL when relevant.

In Fig.~\ref{fig:fluxes}, we show IGRB fluxes for a range of $m_a$ for $\tau_\chi=10^{19}$~sec, with current bounds, and the projected sensitivity of APT. The enveloping shape of the contours follows similar behavior to $\bar{P}^{(1)}_{a\rightarrow\gamma}(z)$:
\begin{enumerate}
    \item \textbf{Fixed $g_{a\gamma\gamma}$ and $m_a$}. As $E_a$ increases, $l_\text{osc}$ increases until $l_\text{osc}\gg \langle l_f\rangle$ and the flux reaches a maximum value. The flux remains close to this maximum value until $l_\text{osc}\ll \langle l_f\rangle$ when $E_a$ is sufficiently large (see Fig.~\ref{fig:losc}), and the flux decreases again (this value of $E_a$ is beyond $10^3$~GeV in the plot shown; in this region, one must include the effects of cascades which would modify the flux profiles, see discussion below). 
    
    \item \textbf{Fixed $g_{a\gamma\gamma}$}. As $E_a$ increases and $m_a$ decreases, the spacing between $m_a$ contours decreases because the oscillation length becomes $m_a$ independent (see Fig.~\ref{fig:dPdzaverage}).
    
    \item \textbf{Fixed $m_a$}. As $g_{a\gamma\gamma}$ increases the flux scales as $g_{a\gamma\gamma}^2$, independent of $E_a$.
\end{enumerate}

Photons produced from axions that originate from decays of dark matter with mass above $1$~TeV will undergo interactions with CMB photons as they traverse cosmological distances. These interactions will result in electromagnetic cascades in which the energy of the initial photons gets distributed to a larger number of photons with lower energies. These lower energies can then fall within the energy range of Fermi-LAT, EGRET, COMPTEL, INTEGRAL, and APT, allowing us to constrain dark matter masses above $1$~TeV~\cite{Blanco_2019}. As in Ref.~\cite{Dunsky:2025pvd}, we expect the strongest constraints on these heavier dark matter candidates to be comparable to TeV mass dark matter up to $\mathcal{O}(1)$ factors. We therefore choose to constrain dark matter masses up to 1~TeV and avoid cascade effects.

An IGRB flux may also be produced directly via 1-loop induced $\chi\rightarrow\gamma\gamma$ decays. This process, however, has a decay lifetime of $\tau_{\chi\rightarrow\gamma\gamma}\sim\tau_\chi/(g_{a\gamma\gamma}m_\chi)^4\gg \tau_\chi$ and contributes a negligible IGRB flux compared to that from axion-photon conversions within filaments. In addition, axions may decay directly to two photons. However, for the parameter space we constrain this is a negligible contribution-- see Appendix~\ref{app:direct_decays}.

\section{Results}\label{sec:lifetimes}

As discussed in the previous section, the IGRB flux in Eq.~(\ref{eg:photon_flux}) is parameterized by $m_\chi$, $\tau_\chi$, $m_a$ and $g_{a\gamma\gamma}$. Observational constraints of the IGRB flux therefore provide a powerful indirect probe connecting dark matter and axion properties. In this section, we present the main results of our work-- limits on the axion mass and axion-photon coupling for benchmark dark matter masses and decay lifetimes.

In Fig.~\ref{fig:constraints}, we plot constraints of the canonical $(m_a,g_{a\gamma\gamma})$ plane for various dark matter lifetimes~\footnote{The excluded region is determined using the interpolated IGRB flux constraints. For each $(m_a,g_{a\gamma\gamma})$, if for any energy the IGRB flux exceeds that of the interpolated constraints, the point is excluded.}. In the left panel, we take $m_\chi=10^3$~GeV and $B_0=250$~nG as an optimistic benchmark. We restrict to lifetimes above the $\Lambda$CDM Large Scale Structure (LSS) constraint of $\tau_\chi\gtrsim10^{19}$~sec~\cite{Poulin_2016}. For a dark matter lifetime that saturates the LSS constraint, shown in green in Fig.~\ref{fig:constraints}, the IGRB flux constraints exclude a significant portion of parameter space, including a portion of the QCD axion band. In this benchmark, our results improve on current bounds by up to seven orders of magnitude. In contrast, the right panel shows our results for a more conservative benchmark with $m_\chi=10$~GeV and $B_0=1$~nG. In this case, our results improve on current bounds by up to four orders of magnitude.

\section{Conclusion}\label{sec:conclusion}
In this work, we have placed new constraints on the CaB assuming this population arises from dark matter decays to axions. Our novel approach leverages axion-photon conversion in cosmological filaments and we constrain the axion mass and photon coupling for a range of dark matter masses and lifetimes. In our analysis, we determine the axion-induced contribution to the IGRB photon flux assuming axion-photon conversions in cosmological filaments. Our results provide conditional but strong constraints for $m_a\lesssim10^{-5}$~eV for dark matter decay lifetimes up to $\tau_\chi\approx10^{30}$ sec.

The constraints computed in this work conservatively assumes cosmological filaments as the only background magnetic field source for axion-photon conversions. Taking into account additional cosmological magnetic field sources, such as voids or clusters (see Appendix~\ref{app:cluster_voids}), these bounds can be strengthened.

\begin{acknowledgments}
We thank David Dunsky, Keisuke Harigaya, and Elena Pinetti for helpful conversations.
This manuscript has been authored in part by Fermi Forward Discovery Group, LLC under Contract No.
89243024CSC000002 with the U.S. Department of Energy, Office of Science, Office of High Energy Physics. We also acknowledge support from the Kavli Institute
for Cosmological Physics at the University of Chicago
through an endowment from the Kavli Foundation and
its founder Fred Kavli
\end{acknowledgments}

\appendix

\section{Models}\label{app:models}
In this Appendix, we discuss some models that can give rise to dark matter decays to axions on a cosmological timescale.

\subsection{Light PQ Fermions}

The role of decaying dark matter can be realized through PQ-symmetry breaking in a technically natural way. Suppose the SM is extended by three fields, one PQ-charged Weyl fermion $\chi$, a neutral Weyl fermion $\eta$, and a PQ-breaking scalar $\Phi$. Enforcing the axial PQ symmetry, we have the fully generic Lagrangian
\begin{align}
    \mathcal L &= \mathcal L_{\rm SM} + \eta^\dag i \bar \sigma_\mu \partial^\mu \eta + \chi^\dag i \bar \sigma_\mu \partial^\mu \chi + |\partial_\mu \Phi|^2 \nonumber \\
    & ~~~~~~~~~~ - (m_0 \eta \eta + y\Phi  \chi \eta + \textrm{h.c.}) -V(\Phi).
\end{align}
Without loss of generality, we can rephase $\Phi$ to make $y$ real. When $\Phi$ gets a vacuum expectation value, it can be written $\Phi = (f_a + r)e^{ia/f_a}$, and the exponential can be removed from the masses by performing the field redefinition $\chi \to e^{-ia(x)/f_a} \chi$, which induces an axion coupling. In the broken phase, the theory contains
\begin{align}
    \mathcal L &\supset \frac{(\partial_\mu a)}{f_a}\chi^\dag \bar \sigma^\mu \chi\nonumber\\ &+ 
    \begin{pmatrix}
        \eta & \chi
    \end{pmatrix}
    \begin{pmatrix}
        m_0 & yf_a/2 \\
        y f_a/2 & 0
    \end{pmatrix}
    \begin{pmatrix}
        \eta \\ \chi
    \end{pmatrix} + \textrm{h.c.}.
\end{align}
Considering the hierarchy $m_0 \gg y f_a$, there is a heavy state of mass $m_H\sim m_0$ and a light state of mass $m_L\sim m_1 \sim y^2 f_a^2/m_0$. The flavor-specific axion coupling then induces a suppressed axion coupling between the two mass eigenstates, proportional to the mixing angle $\theta \sim y f_a/m_0$, 
\begin{align}
    \mathcal L \supset \theta \frac{\partial_\mu a}{f_a} \psi_H^\dag  \bar \sigma^\mu  \psi_L + \textrm{h.c.},
\end{align}
which leads to a decay rate 
\begin{align}
    \Gamma(\psi_H \to \psi_L a) \sim \frac{\theta^2 m_0^3}{f_a^2}.
\end{align}
Notably, $y$ and $m_0$ are both technically natural. If $y$ is set to zero, $U(1)_{PQ}$ splits into two $U(1)$s, one for $\chi$ and one for $\Phi$. When $\eta \to 0$, we also recover a joint $U(1)$ under which $\eta$ and $\chi$ are oppositely charged.

The axion mass in this scenario is unrelated to the mass scale of dark matter. The axion could be the QCD axion, or acquire its mass through another means.

\subsection{Model-Agnostic Scenario}

In this scenario we assume scalar dark matter $\chi$ and decouple the axion mass from the scale of dark matter. We should therefore preserve the shift symmetry of the axion in the operator that generates the decay, which suggests that the leading order operator is
\begin{align}
    \mathcal L \supset \frac{\Lambda}{f_a^2}\chi (\partial_\mu a) (\partial^\mu a) ~. 
\end{align}
This operator breaks the shift symmetry of $\chi$, and therefore whatever physics generates this operator at the scale $\Lambda$ should also generate a mass for $\chi$. 
Therefore, one would expect $m_\chi \sim \Lambda$, and the decay lifetime is
\begin{align}
    \tau_{\chi \to aa} \sim 10^{21}\,\text{sec}  \left(\frac{10^{-18} \, \rm GeV^{-1}}{g_{a\gamma\gamma}}\right)^{4}  \left(\frac{\rm TeV}{m_{\chi}}\right)^{5}.
\end{align}
This model is directly probed by axion-photon conversions for $m_a \lesssim 10^{-8}\,$eV.

\section{Photon Production from Axion Decays}\label{app:direct_decays}
In addition to axion-photon conversions within cosmological filaments, an IGRB flux may be produced from the direct decay of axions to two photons. 

The decay lifetime of an axion to two photons is~\footnote{The lifetime is modified at one-loop by the axion-electron coupling if the axion is electrophilic; however, even if this coupling is non-zero, the additional contribution is highly suppressed compared to the tree level contribution for the parameter space we consider, and is safely neglected.},
\begin{equation}
    \tau_{a\rightarrow \gamma\gamma}\approx10^{74}\text{ s}\left(\frac{10^{-11}\text{ eV}}{m_a}\right)^3\left(\frac{10^{-18}\text{ GeV}^{-1}}{g_{a\gamma\gamma}}\right)^2.
\end{equation}
Even if dark matter decays were to occur early in the universe's history, the axion decay lifetime far exceeds the age of the universe for the parameter space we constrain in this work, and the IGRB flux from these decays will be negligible compared to the flux from axion-photon conversions in cosmological filaments. We therefore safely neglect the axion decay contribution in our calculations.

\section{IGRB flux from Clusters and Voids}\label{app:cluster_voids}
The IGRB flux resulting from axion-photon conversions within an object with magnetic field strength $B$, characteristic size $l$ and volume filling fraction $f_{\text{vol}}$ scales as $B^2 l f_{\text{vol}}$.
In Table~\ref{tab:Blf}, we provide representative values of $B$, $l$ and $f_\text{vol}$ for filaments, clusters and voids. Although the magnetic field strengths of clusters and voids are larger that those of filaments, their relative size and volume filling fraction make them sub-dominant contribution to the IGRB flux, except for filaments with sufficiently weak fields, small size and small volume filling fraction.
\begin{table}[ht!]
\begin{ruledtabular}
\begin{tabular}{ccccc}
Object & $B$ [nG] & $l$ [Mpc] & $f_\text{vol}$ & $B^2 l f_\text{vol}$\\
\colrule
Filaments&$1-100$&$1$&$10\%$&$10^{-1}-10^{3}$\\
Clusters&$10^3$&$10^{-2}$&$1\%$&$10^{2}$\\
Voids&$10^{-7}-10^{-4}$&$1$&$60-80\%$&$10^{-14}-10^{-8}$\\
\end{tabular}
\end{ruledtabular}
\caption{\label{tab:Blf}%
Order of magnitude estimates for the magnetic field strength, effective size and volume filling fraction of filaments, galaxy clusters and voids. Filaments are the dominant contribution to the IGRB flux, unless their field strength is as small as possible.}
\end{table}

\EnableHTMLAmpersand
\bibliography{apssamp.bib}

\end{document}